\documentclass[twocolumn,showpacs,prb,aps,amssymb]{revtex4}

\usepackage{graphicx}
\usepackage{bm}

\begin{document}

\title{Scalar chiral ground states of spin ladders with
four-spin exchanges}

\author{Tsutomu Momoi,$^1$ Toshiya Hikihara,$^2$ Masaaki
Nakamura,$^3$ and Xiao Hu$^2$} \affiliation{$^1$Institute of
Physics, University of Tsukuba,
Tsukuba, Ibaraki 305-8571, Japan\\
$^2$Computational Materials Science Center, National Institute for
Materials Science, Tsukuba, Ibaraki 305-0047, Japan\\
$^3$Department of Applied Physics, Faculty of Science, Tokyo
University of Science, Shinjuku-ku, Tokyo 162-8601, Japan}
\date{Received 19 November 2002}

\begin{abstract}
We show that scalar chiral order can be induced by four-spin
exchanges in the two-leg spin ladder, using the spin-chirality
duality transformation and matrix-product ansatz.
Scalar-chiral-ordered states are found to be exact ground states
in a family of spin ladder models. In this scalar chiral phase,
there is a finite energy gap above the doubly degenerate ground
states and a $Z_2 \times Z_2 \times Z_2$ symmetry is fully broken.
It is also shown that the SU(4)-symmetric model, which is
self-dual under the duality transformation, is on a multicritical
point surrounded by the staggered dimer phase, the staggered
scalar chiral phase, and the gapless phase.
\end{abstract} 

\maketitle

\section{Introduction} Recently, four-spin exchange
interactions have been attracting interest in spin ladder models
and spin-orbital models, because these interactions in fact appear
in many
systems\cite{KugelK,RogerHD,Brehmer,Matsuda,Schmidt,Nunner,%
HondaKW,Coldea} and can induce exotic ground
states.\cite{NersesyanT,KolezhukM,Pati,YamashitaSU2,Azaria,%
Azaria2,Itoi,HikiharaMH,LauchliST} Various types of four-spin
interactions appear associated with diverse mechanisms, e.g.,
cyclic-exchange processes,\cite{Takahashi,SchmidtT,RogerBBCG}
Coulomb repulsion between doubly degenerate orbitals,\cite{KugelK}
and spin-phonon couplings. Though two-leg antiferromagnetic
Heisenberg spin ladders show a rung-singlet ground state and have
an energy gap between the ground state and
excitations,\cite{DagottoR} it was revealed that four-spin
exchanges can induce a gapped staggered dimer (or spin-Peierls)
phase\cite{NersesyanT,KolezhukM,Pati} and a gapless
phase\cite{Pati,YamashitaSU2,Azaria,Azaria2,Itoi} around the
SU(4)-symmetric point and that a gapped phase with a dominant
vector chirality correlation also
appears.\cite{HikiharaMH,LauchliST}

Very recently, L\"auchli, Schmid, and Troyer\cite{LauchliST}
numerically found a new scalar chiral phase in the two-leg
spin-1/2 ladder with four-spin cyclic exchange. Scalar chiral
states, in which both the time-reversal and parity symmetries are
broken, had been discussed in the context of anyon
superconductivity\cite{anyon,anyon2} and the anomalous Hall
effect,\cite{AHE,AHE2} but realization of scalar chiral order in
SU(2)-symmetric systems had been difficult and a challenging
problem. A scalar chiral order due to the four-spin cyclic
exchange was first proposed on the triangular lattice for
magnetism of solid {}$^3$He films,\cite{KuboM,MomoiKN} though
finite-size system analysis could not find evidence for such
ordering, instead showing spin-liquid ground
states.\cite{Misguich1,Misguich2} Study of spin ladders with
four-spin exchanges is expected to clarify the possibility of
exotic magnetism induced by the four-spin interactions.


In this paper, we study the two-leg spin-1/2 ladder with four-spin
exchanges, whose Hamiltonian is given in the next subsection, and
give a rigorous example of a scalar chiral ground state. In our
analysis, the spin-chirality duality transformation we introduced
in Ref.\ \onlinecite{HikiharaMH} plays an important role. A new
class of models that have an exact ground state with scalar chiral
order are constructed by the means of matrix-product ansatz. The
scalar chiral phase has the following nature: (i) the ground
states are doubly degenerate, (ii) there is a finite energy gap
between ground states and excited states, (iii) the ground states
have long-range staggered scalar chiral order and exponentially
decaying spin correlations, and (iv) a $Z_2\times Z_2 \times Z_2$
symmetry is fully broken. It is also found that the phase boundary
of the scalar chiral phase touches the SU(4)-symmetric point and
the SU(4)-symmetric model is on a multicritical point surrounded
by the staggered dimer phase, the staggered scalar chiral phase,
and the gapless phase.

This paper is organized as follows. The model Hamiltonian is given
in the next subsection. In Sec.\ \ref{sec:duality}, we summarize
the duality transformation and its application to the present
model. In Sec.\ \ref{sec:exact_GS}, it is shown that a scalar
chiral state is the exact ground state in a parameter region of
the model Hamiltonian. The nature of the scalar chiral phase is
discussed. In Sec.\ \ref{sec:SU4}, we discuss the phase diagram
around the SU(4)-symmetric point and find that the SU(4) model is
on a multicritical point. Finally in Sec.\ \ref{sec:discuss} we
conclude with a discussion. Appendix contains a unitary
description of the duality transformation.

\subsection*{Model Hamiltonian}\label{sec:Hamiltonian}
The Hamiltonian of the two-leg spin-1/2 ladder with extended
four-spin exchange interactions is defined as
\begin{eqnarray}
{\cal H} &=& J_{\rm l} \sum_l \left( {\bf s}_{1,l} \cdot {\bf
s}_{1,l+1}
                         + {\bf s}_{2,l} \cdot {\bf s}_{2,l+1}
                         \right)
+ J_{\rm r} \sum_l {\bf s}_{1,l} \cdot {\bf s}_{2,l} \nonumber \\
&+& J_{\rm d} \sum_l \left( {\bf s}_{1,l} \cdot {\bf s}_{2,l+1}
                             + {\bf s}_{2,l} \cdot {\bf s}_{1,l+1} \right)
\nonumber \\
&+& J_{\rm rr} \sum_l \left({\bf s}_{1,l  } \cdot {\bf s}_{2,l  }
\right)
                       \left({\bf s}_{1,l+1} \cdot {\bf s}_{2,l+1} \right)
                       \nonumber\\
&+& J_{\rm ll} \sum_l \left({\bf s}_{1,l  } \cdot {\bf s}_{1,l+1}
\right)
        \left({\bf s}_{2,l  } \cdot {\bf s}_{2,l+1} \right)
\nonumber \\
&+& J_{\rm dd} \sum_l \left({\bf s}_{1,l  } \cdot {\bf s}_{2,l+1}
\right)
        \left({\bf s}_{2,l  } \cdot {\bf s}_{1,l+1} \right).
\label{eq:H_ext4}
\end{eqnarray}
This Hamiltonian includes a variety of models: (I) {\it Four-spin
cyclic-exchange model.} When four-spin exchange constants satisfy
$J_{\rm rr}=J_{\rm ll}=-J_{\rm dd}$, the four-spin terms describe
the cyclic-exchange interaction.\cite{RogerHD} (II) {\it SU(2)
$\times$ SU(2) model.} When the parameters satisfy $J_{\rm
r}=J_{\rm d}=J_{\rm rr}=J_{\rm dd}=0$, the Hamiltonian
has an SU(2)$\times$ SU(2) symmetry. This model was studied
extensively as the SU(2) $\times$ SU(2) spin-orbital model. It was
revealed that when $J_{\rm ll}>0$ the ground state has a staggered
dimer (or spin-Peierls) order\cite{NersesyanT,KolezhukM} for $4
J_{\rm l}>J_{\rm ll}$ and is
gapless\cite{YamashitaSU2,Azaria,Azaria2,Itoi} for $-J_{\rm ll}\le
4 J_{\rm l} \le J_{\rm ll}$. (III) {\it SU(4) model.} As a special
case of model II, the Hamiltonian has an SU(4)
symmetry\cite{Pati,YamashitaSU} at $4 J_{\rm l}=J_{\rm ll}$, which
was exactly solved by the Bethe ansatz.\cite{Sutherland}

\section{Duality}\label{sec:duality}
\subsection{Duality transformation} Let us begin with the
spin-chirality duality transformation, which we developed in Ref.\
\onlinecite{HikiharaMH}.
We introduced the duality transformation defining new spin-1/2
pseudospin operators
\begin{eqnarray}
{\bf S}_l &\equiv&
 \frac{1}{2} \left( {\bf s}_{1,l} + {\bf s}_{2,l} \right)
                   - {\bf s}_{1,l} \times {\bf s}_{2,l},
\label{eq:difS} \\
{\bf T}_l &\equiv&
 \frac{1}{2} \left( {\bf s}_{1,l} + {\bf s}_{2,l} \right)
                   + {\bf s}_{1,l} \times {\bf s}_{2,l},
\label{eq:difT}
\end{eqnarray}
which obey the commutation relations of spins and satisfy
$(S^\alpha_l)^2 = (T^\alpha_l)^2 = 1/4$ for $\alpha=x,y$ or $z$.
In the same way, the original spins ${\bf s}_{1,l}$ and ${\bf
s}_{2,l}$ are expressed in terms of ${\bf S}_l$ and ${\bf T}_l$ in
the forms ${\bf s}_{1,l} =\frac{1}{2} \left( {\bf S}_l + {\bf T}_l
\right)
                   + {\bf S}_l \times {\bf T}_l$
and ${\bf s}_{2,l} = \frac{1}{2} \left( {\bf S}_l + {\bf T}_l
\right)
                   - {\bf S}_l \times {\bf T}_l$.
In Appendix, we show that this transformation derives from a
unitary operator $U_{\pi/2}$ in the form ${\bf S}_l=U_{\pi/2}{\bf
s}_{1,l}U_{\pi/2}^\dagger$ and ${\bf T}_l=U_{\pi/2}{\bf
s}_{2,l}U_{\pi/2}^\dagger$. Since the following relations hold,
\begin{eqnarray}
{\bf s}_{1,l} + {\bf s}_{2,l} &=& {\bf S}_l + {\bf T}_l,
\nonumber \\
{\bf s}_{1,l} - {\bf s}_{2,l} &=& 2~ {\bf S}_l \times {\bf T}_l,
\nonumber \\
- 2~ {\bf s}_{1,l} \times {\bf s}_{2,l} &=& {\bf S}_l - {\bf T}_l,
\nonumber
\end{eqnarray}
this transformation exchanges the N\'eel-type spin and vector
chirality degrees of freedom on the same rung. As an example of
the duality, one can show that the transformation of the dimer (or
spin-Peierls) operator
\begin{equation}\label{eq:op_sd}
{\cal O}_{\rm D}(l) = {\bf s}_{1,l} \cdot {\bf s}_{1,l+1}-{\bf
s}_{2,l} \cdot {\bf s}_{2,l+1}
\end{equation}
leads to the scalar chiral operator
\begin{eqnarray}
{\cal O}_{\rm SC}(l)&=& ({\bf s}_{1,l}+{\bf s}_{2,l})
\cdot ({\bf s}_{1,l+1}\times {\bf s}_{2,l+1})\nonumber\\
& &+ ({\bf s}_{1,l} \times {\bf s}_{2,l}) \cdot ({\bf
s}_{1,l+1}+{\bf s}_{2,l+1}).
\end{eqnarray}
It is known that a group of two-leg ladders with four-spin
interactions shows the staggered dimer order in the ground
state.\cite{NersesyanT,KolezhukM,Pati} The above duality relation
hence shows that their dual models have the staggered ``scalar
chiral" order in the ground state. We will discuss these models in
Secs.\ \ref{sec:exact_GS} and \ref{sec:SU4}.

We now consider the transformation of the spin states on rungs.
Since the total spin on each rung is conserved under the
transformation, the fully polarized spin state
$|\uparrow\rangle_{1,l}|\uparrow\rangle_{2,l}$ is transformed to
$|\uparrow\rangle_{S,l}|\uparrow\rangle_{T,l}$, where
$|\alpha\rangle_{\mu,l}$ ($\alpha=\uparrow,\downarrow$ and
$\mu=1,2$) denotes the spin state operated by ${\bf s}_{\mu,l}$,
and $|\alpha\rangle_{S,l}$ ($|\alpha\rangle_{T,l}$) is the
pseudospin state operated by ${\bf S}_l$ (${\bf T}_l$),
respectively. By applying $S_l^-$ and $T_l^-$ to
$|\uparrow\rangle_{S,l}|\uparrow\rangle_{T,l}$, all pseudospin
states in dual space can be constructed in the forms
\begin{eqnarray}
|\uparrow\rangle_{S,l}|\uparrow\rangle_{T,l}&=&
|\uparrow\rangle_{1,l}|\uparrow\rangle_{2,l},\nonumber\\
|\uparrow\rangle_{S,l}|\downarrow\rangle_{T,l} &=&
\frac{e^{-i\pi/4}}{\sqrt{2}}(|\uparrow\rangle_{1,l}|\downarrow\rangle_{2,l}
+i|\downarrow\rangle_{1,l}|\uparrow\rangle_{2,l}),\nonumber\\
|\downarrow\rangle_{S,l}|\uparrow\rangle_{T,l} &=&
\frac{e^{i\pi/4}}{\sqrt{2}}(|\uparrow\rangle_{1,l}|\downarrow\rangle_{2,l}
-i|\downarrow\rangle_{1,l}|\uparrow\rangle_{2,l}),\nonumber\\
|\downarrow\rangle_{S,l}|\downarrow\rangle_{T,l} &=&
|\downarrow\rangle_{1,l}|\downarrow\rangle_{2,l}.\label{eq:duality_states}
\end{eqnarray}
Using eigenstates for the total spin on each rung, one finds that
this transformation corresponds to a gauge transformation of the
singlet bond state $|s\rangle_l\rightarrow -i|s\rangle_l$, while
it keeps the triplet states invariant. See also Appendix for
further arguments.

\subsection{Duality in the model Hamiltonian} We now apply the
transformation to the model (\ref{eq:H_ext4}) and consider the
duality relation in parameter space. Under this duality
transformation, the total form of the Hamiltonian
(\ref{eq:H_ext4}) remains invariant, but the couplings change. The
couplings of the dual Hamiltonian $\tilde{\cal H}$ are given by
\begin{eqnarray}
\tilde{J}_{\rm r}&=& J_{\rm r},~~~~~
\tilde{J}_{\rm rr}=J_{\rm rr},\nonumber\\
\tilde{J}_{\rm l}&=& \frac{1}{2}(J_{\rm l}+J_{\rm d})
+\frac{1}{8}(J_{\rm ll}-J_{\rm dd}),\nonumber\\
\tilde{J}_{\rm d}&=& \frac{1}{2}(J_{\rm l}+J_{\rm d})
-\frac{1}{8}(J_{\rm ll}-J_{\rm dd}),\nonumber\\
\tilde{J}_{\rm ll}&=& 2(J_{\rm l}-J_{\rm d})
+\frac{1}{2}(J_{\rm ll}+J_{\rm dd}),\nonumber\\
\tilde{J}_{\rm dd}&=& -2(J_{\rm l}-J_{\rm d}) + \frac{1}{2}(J_{\rm
ll}+J_{\rm dd}).
\end{eqnarray}
To see the mapping in the parameter space, we rewrite the
Hamiltonian (\ref{eq:H_ext4}) in the form
\begin{eqnarray}
{\cal H} &=& J_{\rm r} \sum_l {\bf s}_{1,l} \cdot {\bf s}_{2,l}
+J_{\rm rr} \sum_l \left({\bf s}_{1,l  } \cdot {\bf s}_{2,l }
\right)
                \left({\bf s}_{1,l+1} \cdot {\bf s}_{2,l+1} \right)
\nonumber \\
&+& W \sum_l \left({\bf s}_{1,l  } + {\bf s}_{2,l } \right) \cdot
\left({\bf s}_{1,l+1} + {\bf s}_{2,l+1} \right) \nonumber \\
&+& X \sum_l \{ \left({\bf s}_{1,l } \cdot {\bf s}_{1,l+1} \right)
         \left({\bf s}_{2,l  } \cdot {\bf s}_{2,l+1} \right)
         \nonumber\\
& &       + \left({\bf s}_{1,l  } \cdot {\bf s}_{2,l+1} \right)
         \left({\bf s}_{1,l+1} \cdot {\bf s}_{2,l  } \right) \}
         \nonumber\\
&+& Y \sum_l \{ \left({\bf s}_{1,l } - {\bf s}_{2,l  } \right)
\cdot
         \left({\bf s}_{1,l+1} - {\bf s}_{2,l+1}
         \right)\nonumber\\
& &     + 4 \left({\bf s}_{1,l  } \times {\bf s}_{2,l  } \right)
\cdot
         \left({\bf s}_{1,l+1} \times {\bf s}_{2,l+1} \right) \}
         \nonumber \\
&+& Z \sum_l \{ \left({\bf s}_{1,l } - {\bf s}_{2,l  } \right)
\cdot
         \left({\bf s}_{1,l+1} - {\bf s}_{2,l+1}
         \right)\nonumber\\
& &     - 4 \left({\bf s}_{1,l  } \times {\bf s}_{2,l  } \right)
\cdot
         \left({\bf s}_{1,l+1} \times {\bf s}_{2,l+1} \right) \},
\label{eq:H_ext4_2}
\end{eqnarray}
where
\begin{eqnarray}
W&=&\frac{1}{2}(J_{\rm l}+J_{\rm d}),~~~~~
X=\frac{1}{2}(J_{\rm ll}+J_{\rm dd}),\nonumber\\
Y&=&\frac{1}{16}(J_{\rm ll}-J_{\rm dd})+\frac{1}{4}(J_{\rm
l}-J_{\rm d}),\nonumber\\
Z&=&-\frac{1}{16}(J_{\rm ll}-J_{\rm dd})+\frac{1}{4}(J_{\rm
l}-J_{\rm d}).
\end{eqnarray}
Straightforward calculations show that the duality transformation
maps the parameters $(J_{\rm r},J_{\rm rr},W,X,Y,Z)$ to $(J_{\rm
r},J_{\rm rr},W,X,Y,-Z)$; the transformation changes only the
coupling $Z$ of the last term to $-Z$, but leaves the other terms
invariant. The Hamiltonian (\ref{eq:H_ext4}) is, thus, self-dual
at the surface defined by $Z=0$, i.e.,
\begin{equation}
J_{\rm ll}-J_{\rm dd}-4(J_{\rm l}-J_{\rm d})=0.
\end{equation}
The parameter space of the Hamiltonian has six dimensions in
total, and the self-dual surface divides the parameter space into
two regions $Z>0$ and $Z<0$. It should be remarked that the
SU(4)-symmetric model ($4J_{\rm l}=J_{\rm ll}$, $J_{\rm r}=J_{\rm
d}=J_{\rm rr}=J_{\rm dd}=0$) exists on the self-dual surface.
Duality around this specific model will be discussed in Sec.\
\ref{sec:SU4}.

\subsection{$U(1)$ symmetry in the self-dual models}
Here we describe a U(1) symmetry in the self-dual models. Consider
a continuous transformation with the following unitary operator
\[
 U_{\theta} = \prod_l \exp\left[ i\theta
 \left({\bm s}_{1,l}\cdot{\bm s}_{2,l}-\frac{1}{4}\right) \right].
\]
This is a continuous extension of the duality transformation (see
Appendix) and it continuously transforms the dimer operator to the
scalar chiral one. One can show that the Hamiltonian is invariant
under this transformation as $U_\theta {\cal H}
U_\theta^{\dagger}={\cal H}$ for arbitrary $\theta$ if $Z=0$. Thus
the self-dual models are isotropic under the continuous rotation
with the generator $\sum_l {\bf s}_{1,l}\cdot {\bf s}_{2,l}$,
whereas the $Z$-term of the Hamiltonian (\ref{eq:H_ext4_2}) lowers
the symmetry.

\section{Models with exact scalar chiral ground states}
\label{sec:exact_GS} In this section, we discuss an exact scalar
chiral ground state of the Hamiltonian (\ref{eq:H_ext4}) with the
periodic boundary condition. To obtain the ground state $\Psi_0$,
we use matrix-product (MP) states. We start from the following
{\it ansatz}:
\begin{equation}
\Psi_0(u) = {\rm tr}\{ \tilde{\bm g}_1(u) \tilde{\bm g}_2(-u)
\cdots \tilde{\bm g}_{2N-1}(u) \tilde{\bm g}_{2N}(-u) \},
\label{eq:MP_ansatz}\end{equation} where $u$ is a real variable
and
\begin{eqnarray}
& &\tilde{\bm g}_l (u) = \frac{1}{2} \left( \begin{array}{cc}
  i u|s\rangle_l+|t_0\rangle_l~~ & -\sqrt{2}|t_{+1}\rangle_l \\
  \sqrt{2}|t_{-1}\rangle_l~~ & i u|s\rangle_l-|t_0\rangle_l
\end{array}\right)\\
&&= \frac{1}{2} \{i u{\bm 1}|s\rangle_l -\sqrt{2}\sigma^+
|t_{+1}\rangle_l + \sqrt{2}\sigma^- |t_{-1}\rangle_l + \sigma_z
|t_0\rangle_l\}.\nonumber
\end{eqnarray}
Here $|s\rangle_l$ and $|t_\mu \rangle_l$ are, respectively, the
singlet and triplet states of the $l$th rung, $2N$ is the total
number of rungs, $\bm 1$ is the $2\times 2$ unit matrix, and
$\sigma_\mu$ are the Pauli matrices. This form of the MP state can
be obtained by the duality transformation of the MP state
discussed by Kolezhuk and Mikeska,\cite{KolezhukM}
\begin{equation}
\Psi_{\rm KM}(u)={\rm tr}\{ {\bm g}_1(u) {\bm g}_2(-u) \cdots {\bm
g}_{2N-1}(u) {\bm g}_{2N}(-u) \},
\end{equation}
where
\begin{equation}
{\bm g}_l (u) = \frac{1}{2} \{u{\bm 1}|s\rangle_l
-\sqrt{2}\sigma^+ |t_{+1}\rangle_l + \sqrt{2}\sigma^-
|t_{-1}\rangle_l + \sigma_z |t_0\rangle_l\}.
\end{equation}
It was shown that for several models this MP state $\Psi_{\rm
KM}(u)$ is the exact ground state with a staggered dimer order. At
$u=0$ and $u=\infty$, the two states $\Psi_0(u)$ and $\Psi_{\rm
KM}(u)$ are equivalent, which means that each of $\Psi_0(0)$ and
$\Psi_0(\infty)$ is self-dual. $\Psi_0(0)$ and $\Psi_0(\infty)$
are, respectively, an Affleck-Kennedy-Lieb-Tasaki (AKLT)
state\cite{AKLT} and a rung-singlet state. For $0<u<\infty$,
however, $\Psi_0(u)$ and $\Psi_{\rm KM}(u)$ are orthogonal in the
limit $N\rightarrow\infty$.

Since $\Psi_{\rm KM} (u)$ ($0<u<\infty$) has the staggered dimer
order, $\Psi_0(u)$ has the staggered scalar chiral order because
of the duality relation. Using the technique developed by
Kl\"umper, Schadschneider, and Zittartz,\cite{KlumperSZ,Fannes}
one can evaluate the scalar chiral correlation in $\Psi_0(u)$,
\begin{equation}\label{eq:s_chi_corr}
\langle {\cal O}_{\rm SC}(l) {\cal O}_{\rm
SC}(m)\rangle=(-1)^{l-m}\left[\frac{12u}{(u^2+3)^2}\right]^2,
\end{equation}
for $|l-m|> 1$ in the limit $N\rightarrow\infty$ and show
spontaneous breakdown of the chiral symmetry
\[
\langle {\cal O}_{\rm SC}(l) \rangle =
(-1)^{l}\frac{12u}{(u^2+3)^2}.
\]
One can also evaluate that $\Psi_0(u)$ has no dimer correlation
\[
\langle {\cal O}_{\rm D}(l) {\cal O}_{\rm D}(m)\rangle = 0.
\]
In the same way, the spin and vector chiral correlations are
obtained as
\begin{eqnarray}
\lefteqn{\langle s_{1,l}^\alpha s_{1,m}^\alpha \rangle=
\langle s_{1,l}^\alpha s_{2,m}^\alpha \rangle}\nonumber\\
& &~~~=
-\frac{1}{(u^2+3)(u^2-1)}\left(\frac{u^2-1}{u^2+3}\right)^{|l-m|},
\label{eq:spin_corr}\\
\lefteqn{\langle ({\bf s}_{1,l}\times {\bf s}_{2,l})^\alpha
({\bf s}_{1,m}\times {\bf s}_{2,m})^\alpha \rangle} \nonumber\\
& &~~~=
\frac{u^{2}}{(u^2+3)(u^2-1)}\left(\frac{1-u^2}{u^2+3}\right)^{|l-m|},
\label{eq:v_chi_corr}
\end{eqnarray}
for $\alpha=x,$ $y$ or $z$. The spin and vector chiral correlation
lengths are equal and given by $\xi_{\rm s}^{-1}=\xi_{\rm
vc}^{-1}=\ln\{(u^2+3)/|u^2-1|\}$, whereas the scalar chiral
correlation does not have any exponentially decaying term.

By the duality transformation of the MP-solvable models presented
by Kolezhuk and Mikeska,\cite{KolezhukM} we find that the MP state
(\ref{eq:MP_ansatz}) is an exact ground state of the following
three classes of models. One can prove that the state $\Psi_0$ is
a ground state in these models, reducing a local Hamiltonian
$h_{l,l+1}$ on the $l$th and $l+1$th rungs to a positive
semi-definite form $(h_{l,l+1}-E_0)\ge 0$, where ${\cal H}=\sum_l
h_{l,l+1}$, and showing that $\Psi_0$ has zero eigenenergy in the
reduced Hamiltonian as $(h_{l,l+1}-E_0){\bm g}_l {\bm g}_{l+1}$=0.
The reader who is interested in the method of proofs should refer
to Ref.\ \onlinecite{KolezhukM}.

(A) {\it Scalar chiral models.} For a family of models
\begin{eqnarray}
J_{\rm r}&=&\frac{8J(2-3y)}{3(4-3y)},~~~~~
J_{\rm l}=\frac{4J(1-y)}{4-3y}, \nonumber\\
J_{\rm d}&=&\frac{J(8-9y)}{3(4-3y)},~~~~~
J_{\rm ll}=\frac{16J}{3(4-3y)}, \nonumber\\
J_{\rm rr}&=& 0, ~~~~~~~~ J_{\rm dd}=
\frac{-4Jy}{4-3y},\label{eq:scalar}
\end{eqnarray}
with $0< y < 1$ and $J>0$, the ground states are doubly degenerate
and given by $\Psi_0(1)$ and $\Psi_0(-1)$. The ground-state energy
per rung is $E_0=-3J/4$. This model is dual to the
``checkerboard-dimer model" given in Ref.\ \onlinecite{KolezhukM},
which has a staggered dimer order, and hence from the duality
relation the present model belongs to the scalar chiral phase.
Excitations of the staggered dimer phase were studied by
variational trial states,\cite{KolezhukM} numerical
calculations,\cite{Pati} and field-theoretical
analyses.\cite{Azaria,Azaria2,Itoi} These studies concluded that
excitations have a finite energy gap. In fact, extending the
arguments by Knabe,\cite{Knabe,note} we can prove the finiteness
of the energy gap in the checkerboard-dimer model dual to the
model (\ref{eq:scalar}) with $y=2/3$ and in a finite region around
this point.  From the duality we conclude that, in the scalar
chiral phase, there is a finite energy gap between the ground
states and excited states.
The analysis in the dual model\cite{KolezhukM} indicates that, at
$y=1$, the present system enters the fully polarized ferromagnetic
phase through a first-order transition. Furthermore, we can extend
the parameter space which has the exact scalar chiral ground
state. For $y=2/3$, the dual Hamiltonian has an SU(2) $\times$
SU(2) symmetry and the Hamiltonian is written as a product of
projection operators $(4J/3)\sum_{l}({\bf s}_{1,l}\cdot{\bf
s}_{1,l+1}+3/4)({\bf s}_{2,l}\cdot{\bf s}_{2,l+1}+3/4)$. Then one
can construct the model with the exact scalar chiral ground state
by generalization of projection operators\cite{Itoh} and the
duality transformation.

(B) {\it Model 
at a phase boundary between the scalar chiral and staggered dimer
phases.} At $y=0$ of the model (\ref{eq:scalar}), i.e.,
\begin{eqnarray}
J_{\rm r}&=& J_{\rm ll}=\frac{4J}{3},~~~~~~
J_{\rm l}= J, \nonumber\\
J_{\rm d}&=&\frac{2J}{3},~~~~~~ J_{\rm rr}= J_{\rm dd}= 0,
\label{eq:solvable2}
\end{eqnarray}
with $J>0$, the ground states are given by $\Psi_0(u)$ with
arbitrary $u$ and highly degenerate. Note that this model is
equivalent to the ``multicritical model" in Ref.\
\onlinecite{KolezhukM} and self-dual under the duality
transformation. The scalar chiral model (\ref{eq:scalar}), thus,
connects with the checkerboard-dimer model at this special
parameter point.
At this phase boundary, both one magnon and a pair of scattering
solitons have energy gaps in the staggered dimer state, as
discussed by Kolezhuk and Mikeska.\cite{KolezhukM} However,
because of the U(1) symmetry in this model, the generator ${\bf
s}_{1,l}\cdot {\bf s}_{2,l}$ can create gapless collective
(Goldstone) modes, which are singlet bound states of two magnons.
In fact, one can show that the following trial state becomes
gapless at $p=0$, $\pi$ ($\zeta=-1$) and $p=\pi/2$ ($\zeta=1$):
\begin{eqnarray}
|\Psi_0 (p)\rangle_{sb}^\zeta &=& \sum_l e^{2ipl} {\rm Tr}\left\{
\prod_{i=1}^{l-1}\tilde{\bm g}_{2i-1}(1) \tilde{\bm g}_{2i} (-1)
{\bm g}_{l}^{sb,\zeta}(1) \right.\nonumber\\
&\times& \left. \tilde{\bm g}_{2l+2}
(-1)\prod_{i=l+2}^{N}\tilde{\bm
g}_{2i-1}(1) \tilde{\bm g}_{2i}(-1) \right\},\nonumber\\
{\bm g}_{l}^{sb,\zeta}(1) &=& \left\{\sum_\alpha \sigma^\alpha
\tilde{\bm g}_{2l-1}(1) \sigma^\alpha \tilde{\bm g}_{2l} (-1)
\right\} \tilde{\bm g}_{2l+1} (1) \nonumber\\
&+& \zeta \tilde{\bm g}_{2l-1}(1) \left\{\sum_\alpha \sigma^\alpha
\tilde{\bm g}_{2l}(-1) \sigma^\alpha \tilde{\bm g}_{2l+1}
(1)\right\}. \nonumber
\end{eqnarray}

(C) {\it Model with two second-order phase boundaries.} For a
family of models
\begin{eqnarray}
J_{\rm r}&=&-J_{\rm rr}=\frac{J}{6}(u^2-1)(u^2+3),\nonumber\\
J_{\rm l}&=&\frac{J}{48}(3u^2+5)(u^2+3),~~~~~
J_{\rm d}=\frac{J}{3}u^2,\label{eq:model_c}\\
J_{\rm ll}&=&\frac{J}{12}(5u^2+3)(u^2+3),~~~~~ J_{\rm
dd}=\frac{J}{6}(u^4-6u^2-3),\nonumber
\end{eqnarray}
with arbitrary $u$, the ground states are $\Psi_0(u)$ and
$\Psi_0(-u)$, and the ground-state energy per rung is
$E_0=-J(7u^4+22u^2+19)/64$. The model at $u=1$ is equivalent to
the model (\ref{eq:scalar}) at $y=2/3$.
The arguments for the dual model\cite{KolezhukM} lead to the
conclusion that the model (\ref{eq:model_c}) undergoes a phase
transition to the Haldane phase at $u=0$ and to the rung-singlet
phase at $u=\infty$. Both of the transitions are of second order
and accompanied with vanishing of energy gaps for solitons.

\begin{figure}[tb]
    \includegraphics[width=80mm]{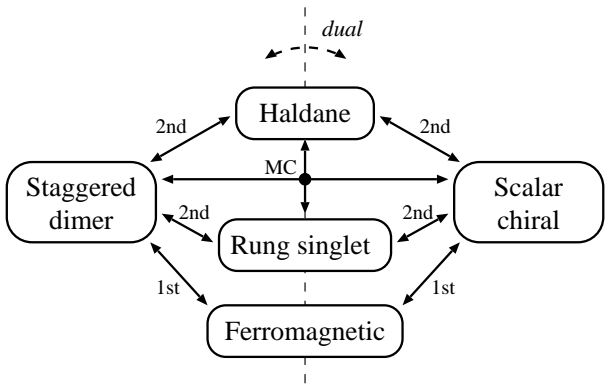}
\caption{Schematic picture of five phases and phase transitions
that appear in MP-solvable models. Phase transitions occur along
arrows. The number attached to each arrow denotes the order of the
phase transition. Black circle (MC) denotes the multicritical
point (\ref{eq:solvable2}). The scalar chiral phase is dual to the
staggered dimer phase, while Haldane, rung-singlet, and
ferromagnetic phases are self-dual.} \label{fig:pd_exact}
\end{figure}

In total, five phases appear in the MP-solvable models discussed
above and in Ref.\ \onlinecite{KolezhukM}. Some of the phase
transitions between them actually happen in the parameter space of
the solvable models. These are summarized in Fig.\
\ref{fig:pd_exact}.

The nature of the scalar chiral phase is summarized as follows:
(1) the ground states are doubly degenerate, (2) there is a finite
energy gap between the ground states and excited states, and (3)
the ground states have long-range staggered scalar chiral order
and exponentially decaying spin correlations.

It is also easy to show that the string order, which was
originally found in the Haldane state,\cite{dNijsR,GirvinA} exists
in the scalar chiral state $\Psi_0(u)$. The expectation value is
given by
\begin{eqnarray}
& &\langle (s_{1,j}^\alpha+s_{2,j}^\alpha)
\prod_{l=j}^{k-1}\exp\{i\pi(s_{1,l}^\alpha+s_{2,l}^\alpha)\}
(s_{1,k}^\alpha+s_{2,k}^\alpha)\rangle \nonumber\\
& &~~~~~=4/(u^2+3)^2 \label{eq:str_corr}
\end{eqnarray}
for $\alpha=x,$ $y$, or $z$. We note that the staggered dimer
state $\Psi_{\rm KM}(u)$ also has exactly the same expectation
value of the string order because the string operator is invariant
under the duality transformation. This string order implies that a
hidden $Z_2 \times Z_2$ symmetry is spontaneously broken in the
ground state.\cite{KennedyT} This hidden symmetry was
found\cite{KennedyT,TakadaW} by applying a nonlocal unitary
transformation $U$, and one can find that this symmetry exists
also in the Hamiltonian (\ref{eq:H_ext4}). It should be noted that
this $Z_2 \times Z_2$ symmetry is independent of the $Z_2$ chiral
symmetry associated with the scalar chirality, since the scalar
chiral operator after the nonlocal unitary transformation $U{\cal
O}_{\rm SC}U^{-1}$ has the corresponding $Z_2 \times Z_2$
symmetry. In finite systems with open boundary conditions, the
scalar chiral MP ground states in fact
have eightfold degeneracy associated with boundary spins and
chirality. Thus, the hidden $Z_2 \times Z_2$ symmetry, as well as
the $Z_2$ chiral symmetry, is spontaneously broken in the scalar
chiral phase. Recently, a different useful quantity
$z_{2N}=\langle \exp[(2\pi i /2N) \sum_{l=1}^{2N}
l(s_{1,l}^z+s_{2,l}^z)]\rangle$ was proposed,\cite{Nakamura} which
detects the average number $n$ of valence bonds between
neighboring rungs as $\lim_{N\rightarrow\infty}z_{2N}=(-1)^n$. In
$\Psi_0(u)$, the expectation value is estimated as
$\lim_{N\rightarrow\infty} z_{2N} = -1$ for finite $u$. This is
consistent with the above valence bond picture, because $z_{2N}$
is also invariant under the duality transformation, and $n=1$ in
the staggered dimer state.

\section{Around the SU(4)-symmetric point}\label{sec:SU4}
Using the duality relation, we discuss the phase diagram around
the SU(4)-symmetric point and show that this SU(4)-symmetric point
is a multicritical point.

\subsection{SU(2) $\times$ SU(2) spin ladders}
We start from two SU(2) $\times$ SU(2) spin ladders. One SU(2)
$\times$ SU(2) spin ladder is model II. Here we consider the case
$J_{\rm l}\ge 0$.
For $J_{\rm l}/J_{\rm ll} = 1/4$, this model is SU(4) symmetric
and exactly solvable by the Bethe ansatz, and the ground state is
gapless critical.\cite{Sutherland} For $J_{\rm l}/J_{\rm ll} >
1/4$, the ground state has a finite gap and a staggered dimer
order,\cite{KolezhukM,Pati,YamashitaSU2} whereas for $0\le J_{\rm
l}/J_{\rm ll} \le 1/4$ the ground state is gapless and
critical.\cite{Pati,YamashitaSU2,Azaria,Azaria2,Itoi} The total
Hamiltonian can be divided into the part of the SU(4) model ${\cal
H}_0$ ($J_{\rm ll}=4J_{\rm l}$ with fixed $J_{\rm l}$) and the
perturbation ${\cal V}$ in the form
\begin{eqnarray}
{\cal H}^\prime&=&{\cal H}_0+\lambda{\cal V},
\label{eq:su2xsu2}\\
{\cal H}_0 &=& J_{\rm l} \sum_l \{\left( {\bf s}_{1,l} \cdot {\bf
s}_{1,l+1}
                         + {\bf s}_{2,l} \cdot {\bf s}_{2,l+1}
                         \right)\nonumber\\
& &+ 4 \left({\bf s}_{1,l  } \cdot {\bf s}_{1,l+1} \right)
        \left({\bf s}_{2,l  } \cdot {\bf s}_{2,l+1}
        \right)\},\nonumber\\
{\cal V} &=& \sum_l \left({\bf s}_{1,l  } \cdot {\bf s}_{1,l+1}
\right)
        \left({\bf s}_{2,l  } \cdot {\bf s}_{2,l+1} \right),
\nonumber
\end{eqnarray}
where $\lambda\equiv -4J_{\rm l }+J_{\rm ll}$ and $J_{\rm l}$ is
fixed. Renormalization group analysis concluded that if the
parameter $\lambda$ is negative, the perturbation ${\cal V}$ is
relevant and leads to a generation of a staggered dimer order with
a finite spin gap, and if the coupling parameter is positive, this
perturbation is irrelevant and keeps the ground state
gapless.\cite{Azaria,Azaria2,Itoi}

Applying the spin-chirality duality transformation to Eq.\
(\ref{eq:su2xsu2}), one obtains the dual Hamiltonian
\begin{eqnarray}
\tilde{\cal H}^\prime &=& {\cal H}_0+\lambda
\tilde{\cal V},\label{eq:su2xsu2_dual}\\
\tilde{\cal V} &=& \frac{1}{8} \sum_l
     \left( {\bf s}_{1,l} \cdot {\bf s}_{1,l+1}
          + {\bf s}_{2,l} \cdot {\bf s}_{2,l+1} \right) \nonumber \\
&-& \frac{1}{8} \sum_l
     \left( {\bf s}_{1,l} \cdot {\bf s}_{2,l+1}
          + {\bf s}_{2,l} \cdot {\bf s}_{1,l+1} \right) \nonumber \\
&+& \frac{1}{2} \sum_l
     \left( {\bf s}_{1,l} \cdot {\bf s}_{1,l+1} \right)
     \left( {\bf s}_{2,l} \cdot {\bf s}_{2,l+1} \right) \nonumber \\
&+& \frac{1}{2} \sum_l
     \left( {\bf s}_{1,l} \cdot {\bf s}_{2,l+1} \right)
     \left( {\bf s}_{2,l} \cdot {\bf s}_{1,l+1} \right).\nonumber
\end{eqnarray}
Here ${\cal H}_0$ is self-dual, and $\tilde{\cal V}$ is the
perturbation dual to ${\cal V}$.
The couplings of $\tilde{\cal H}^\prime$, in total, are given as
\begin{eqnarray}
\tilde{J}_{\rm r}&=& \tilde{J}_{\rm rr} = 0,\nonumber\\
\tilde{J}_{\rm l}&=& \frac{1}{2}J_{\rm l} +\frac{1}{8}J_{\rm
ll},~~~~~
\tilde{J}_{\rm d} = \frac{1}{2}J_{\rm l}
-\frac{1}{8}J_{\rm ll},\nonumber\\
\tilde{J}_{\rm ll}&=& 2J_{\rm l}+\frac{1}{2}J_{\rm ll},~~~~~
\tilde{J}_{\rm dd} = -2J_l+\frac{1}{2}J_{\rm ll}.
\end{eqnarray}
From this transformation, one finds that the Hamiltonian
(\ref{eq:su2xsu2_dual}) has a hidden SU(2) $\times$ SU(2)
symmetry, where generators are given by $\sum_l S_l^\alpha$ and
$\sum_l T_l^\alpha$ for $\alpha=x$, $y$, or $z$.
The duality transformation leads to the case where, if the
coupling parameter $\lambda$ is negative, the perturbation
$\tilde{\cal V}$ is relevant and induces a staggered scalar chiral
order with a finite spin gap, and if the coupling parameter is
positive, this perturbation is irrelevant and keeps the ground
state gapless. When $\lambda=-\frac{8}{3} J_{\rm l}$ (i.e.,
$J_{\rm ll}=\frac{4}{3}J_{\rm l}$), the model $\tilde{\cal
H}^\prime$ equals to the scalar chiral model (\ref{eq:scalar})
with $y=2/3$ and as we have shown in Sec.\ III the exact ground
state has an energy gap and the scalar chiral order.

\subsection{Phase diagram around the SU(4) point}
\begin{figure}[tb]
    \includegraphics[width=60mm]{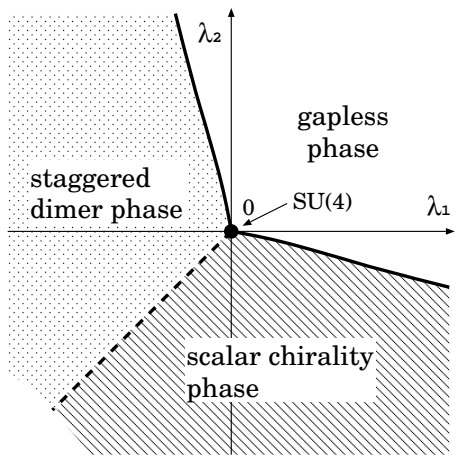}
\caption{Schematic possible phase diagram around the
SU(4)-symmetric point. The phase transition on the solid line is
of second order and that on the dashed line is either first order
or second order. Other phases might be inserted around the phase
boundaries.} \label{fig:pd_su4}
\end{figure}
We now discuss the phase diagram around the SU(4)-symmetric point,
and the two SU(2) $\times$ SU(2) models given above, considering
the following generalized Hamiltonian:
\begin{equation}
{\cal H}^{\prime\prime} = {\cal H}_0 + \lambda_1{\cal V} +
\lambda_2 \tilde{\cal V}.
\end{equation}
This Hamiltonian contains two kinds of perturbation to the
SU(4)-symmetric model. Because of the duality, phase boundaries
are symmetric with the line $\lambda_1=\lambda_2$. The nature of
phases in $\lambda_1>\lambda_2$ is related to that in
$\lambda_1<\lambda_2$ by the duality transformation. The above
consideration leads to a conclusion that the SU(4)-symmetric point
is a multicritical point and surrounded by the staggered dimer
phase, the staggered scalar chiral one, and the critical one.
If the scalar chiral phase touches with the staggered dimer phase,
the phase boundary between two phases must exist exactly on the
self-dual line $\lambda_1=\lambda_2$ (see Fig.\ \ref{fig:pd_su4}).
Because of the U(1) symmetry, a rigorous theorem\cite{Momoi}
concludes that in general
both orders disappear on the self-dual line and hence the
transition between the scalar chiral and dimer phases is second
order, but, if uniform susceptibility of ${\bf s}_{1,l}\cdot{\bf
s}_{2,l}$ diverges, both orders can exist on this line. Note that
the latter actually happens in the model (\ref{eq:solvable2}). One
plausible phase diagram is shown in Fig.\ \ref{fig:pd_su4}.
Recently we have numerically studied this phase diagram and
obtained results consistent with the present
conclusion.\cite{HikiharaNM}

\section{Discussion}\label{sec:discuss}

In this paper we have shown a rigorous example of scalar chiral
ground states in SU(2) spin ladders with four-spin exchanges. The
exact duality relation is the keystone of our theory. Our results
demonstrated that four-spin exchanges can actually induce the
scalar chiral long-range order. The scalar chiral phase extends to
a wide parameter region and touches with the SU(4)-symmetric
point.
Previously, a scalar chiral phase was numerically found in the
four-spin cyclic exchange model on the two-leg
ladder.\cite{LauchliST} In their phase diagram, the scalar chiral
phase appears next to the staggered dimer phase and the phase
boundary indeed exists on the self-dual point\cite{HikiharaMH}
$J_4/J=1/2$. This situation in the vicinity of the self-dual point
shows a resemblance to that around the self-dual line
($\lambda_1=\lambda_2<0$) in Fig.\ \ref{fig:pd_su4}. Our recent
numerical study of the Hamiltonian (\ref{eq:H_ext4}) indicates
that the scalar chiral phase we found in this paper extends to the
four-spin cyclic-exchange case and that two phases belong to the
same one.

We have shown that the SU(4)-symmetric model is self-dual under
the spin-chirality duality transformation. We here note that this
statement holds for the SU(4) spin-orbital models on arbitrary
lattices. Recently SU(4) spin-orbital models on two-dimensional
lattices\cite{Li} and on ladders\cite{Bossche} have been studied
and it was discussed that a plaquette ordering may appear in the
ground state. On a four-site plaquette, the SU(4) singlet state is
the unique ground state and therefore it must be self-dual under
the duality transformation. We hence conclude that plaquette
ordering is also self-dual.

Last, we discuss the universality classes of phase transitions.
The phase transitions into the scalar chiral phase are naturally
in the same universality class as the dual transitions into the
staggered dimer phase. For example, since the phase transition
between the rung-singlet phase and the staggered dimer phase
belongs to the $c=3/2$ SU(2)$_2$ criticality,\cite{NersesyanT} we
conclude that the transition between the scalar chiral phase and
the rung-singlet phase also belongs to the same one. Since the
two-dimensional Ising model is related to the $c=1/2$ criticality,
this $c=3/2$ criticality can be plausibly regarded as a
consequence of the $Z_2 \times Z_2 \times Z_2$ symmetry breaking.

\acknowledgments We would like to thank K.\ Kubo and H.\
Tsunetsugu for stimulating discussions and L.\ Balents for useful
comments. This work was supported by the Ministry of Education,
Culture, Sports, Science and Technology (MEXT) of Japan through
Grants-in-Aid Nos.\ 13740201, 1540362 (T.M.), and No.\ 14740241
(M.N.). T.M. acknowledges kind hospitality of Yukawa Institute in
Kyoto University, where this research was partially performed.

\appendix

\section*{Appendix: Unitary operator for duality transformation}
In this appendix, we show that the duality transformation
(\ref{eq:difS}) and (\ref{eq:difT}) corresponds to a unitary
transformation of two spins on rungs.
The unitary operator is given by
\begin{eqnarray}
 U_{\theta}&=&\prod_l \exp\left[ -i\theta P_l ({\rm s}) \right]
 \nonumber\\
 &=& \prod_l \exp\left[ i\theta
 \left({\bm s}_{1,l}\cdot{\bm s}_{2,l}-\frac{1}{4}\right) \right]
  \label{eq:unit_spin}
\end{eqnarray}
with $\theta=\pi/2$, where $P_l({\rm s})$ denotes the projection
operator onto the singlet state on the $l$th rung. Since the
generator is transformed as ${\bm s}_{1,l}\cdot{\bm s}_{2,l} =
({\bm s}_{1,l}+{\bm s}_{2,l})^2 /2 - 3/4$, this unitary conserves
the total spin on each rung. Note that the generator of this
unitary is a summation of SU(4) generators $s_1^\alpha s_2^\alpha$
($\alpha=x,y,z$) and hence the SU(4)-symmetric model is naturally
invariant under this transformation.

Let us demonstrate the unitary transformation of spins. It is
convenient to reduce the unitary operator to the form
\begin{eqnarray}
 U_{\theta} 
 = \prod_l \left[1+(1-e^{-i\theta})
 \left({\bm s}_{1,l}\cdot{\bm s}_{2,l}-\frac{1}{4}\right)\right],
\end{eqnarray}
where we have used the relation $[P_l({\rm s})]^2=P_l({\rm s})$.
Using the following commutation relations
\begin{eqnarray}
 [{\bm s}_{1,l}\cdot{\bm s}_{2,l}, {\bm s}_{1,l}+{\bm s}_{2,l}] &=& 0,\\
{} [{\bm s}_{1,l}\cdot{\bm s}_{2,l}, {\bm s}_{1,l}-{\bm s}_{2,l}]
 &=& 2 i{\bm s}_{1,l}\times{\bm s}_{2,l},\\
{} [{\bm s}_{1,l}\cdot{\bm s}_{2,l}, {\bm s}_{1,l}\times{\bm
s}_{2,l}] &=& -\frac{i}{2}({\bm s}_{1,l}-{\bm s}_{2,l}),
 \label{eq:spin_commutations}
\end{eqnarray}
and the unitary relation $U_\theta {U_\theta}^\dagger=1$, one can
perform the unitary transformation of spins in the forms
\begin{eqnarray}
 \lefteqn{U_{\theta}({\bm s}_{1,l}+{\bm s}_{2,l}){U_{\theta}}^{\dagger}
  ={\bm s}_{1,l}+{\bm s}_{2,l},}\\
 \lefteqn{U_{\theta}({\bm s}_{1,l}-{\bm s}_{2,l}){U_{\theta}}^{\dagger}}
 \nonumber\\
  &&=\cos\theta({\bm s}_{1,l}-{\bm s}_{2,l})
  -2\sin\theta({\bm s}_{1,l}\times{\bm s}_{2,l}),\\
 \lefteqn{U_{\theta}({\bm s}_{1,l}\times{\bm s}_{2,l}){U_{\theta}}^{\dagger}}
 \nonumber\\
  &&=\frac{1}{2}\sin\theta({\bm s}_{1,l}-{\bm s}_{2,l})
  +\cos\theta({\bm s}_{1,l}\times{\bm s}_{2,l}).
\end{eqnarray}
When $\theta=\pi/2$, we obtain the original duality transformation
\begin{eqnarray}
 {\bm S}_l &=& U_{\pi/2}{\bm s}_{1,l}{U_{\pi/2}}^{\dagger},\\
 {\bm T}_l &=& U_{\pi/2}{\bm s}_{2,l}{U_{\pi/2}}^{\dagger}.
\end{eqnarray}
From the form of the unitary operator, it is clear that this
unitary corresponds to a gauge transformation of the singlet bond
state
\begin{eqnarray}
 U_\theta |s\rangle_{l} &=& e^{-i\theta}|s\rangle_{l},\\
 U_\theta |t_m\rangle_{l} &=& |t_m\rangle_{l}, ~~~~~(m=-1,0, 1).
\end{eqnarray}
When $\theta=\pi/2$, one obtains the relation
(\ref{eq:duality_states}) from
\begin{equation}
 |\sigma\rangle_{S,l} |\sigma^\prime\rangle_{T,l} =
 U_{\pi/2} |\sigma\rangle_{1,l} |\sigma^\prime\rangle_{2,l}
\end{equation}
for $\sigma$ ($\sigma^\prime$) $=\uparrow,\downarrow$.




\begin{thebibliography}{99}
\bibitem{note} For the model (\ref{eq:scalar}) with $y=2/3$, the
local Hamiltonian between the nearest neighbor rungs can be
written as a projection operator if one sets $J=3/4$ and hence one
can apply the rigorous argument in Ref.\ \onlinecite{Knabe} to the
present system. We thereby prove that the energy gap in the
thermodynamic limit is larger than $0.115474J$, where we have used
the value of the energy gap $\Delta \epsilon$ for a finite system
with eight rungs, evaluated as $\Delta \epsilon=0.289454J$.
\end{thebibliography}
\end{document}